\documentclass[11pt]{article}

\usepackage[a4paper,margin=2.5cm]{geometry}
\usepackage{graphicx}
\usepackage{amsmath,amssymb}
\usepackage{hyperref}
\usepackage{authblk}

\title{Studies on photon-feedback and LaB$_6$ photocathode for the GasPM development}

\author[1]{S.~Garnero}
\author[2,3,5]{K.~Inami}
\author[2,3,4,5]{K.~Matsuoka}
\author[1]{R.~Okubo}
\author[2]{K.~Ueda}

\affil[1]{INFN, Sezione di Trieste, Trieste, Italy}
\affil[2]{Nagoya University, Nagoya, Japan}
\affil[3]{KEK, Tsukuba, Japan}
\affil[4]{SOKENDAI, Hayama, Japan}
\affil[5]{Kobayashi-Maskawa Institute (KMI), Nagoya, Japan}

\date{}

\begin{document}
\maketitle

\begin{abstract}
We present new developments, based on beam tests and cosmic rays, on the gaseous photomultiplier (GasPM). The GasPM detects photons by combining a photocathode with a resistive-plate-chamber avalanche. It achieves $\mathcal{O}$(10) ps time resolution with affordable scalability. The  GasPM provides precise and efficient Cherenkov-based charged-particle identification too when combined with a radiator. Our target application  in a future Belle II upgrade aims at suppressing beam-induced background photons, which are typically detected off-collision time, that spoil the electromagnetic calorimeter performance. We reached 25 ps single-photon time-resolution at 3.3 × 10$^6$ gain in 2022, using a picosecond-pulse laser and a LaB$_6$ photocathode. However, electrons entering through a MgF$_2$ window upstream of a CsI photocathode showed a worsening to 70 ps in a 2023 test. Here we aim at addressing the chief causes of the observed degradation. We focus on ultraviolet-photon emission from the de-excitation of the gas molecules, which generates a secondary "photon-feedback" signal overlapping the primary one, and degrading time resolution. We conceive and operate an improved beam test that, along with multiple device-configuration changes, employes a new 10 GSPS frequency digitizer to separate the photon-feedback signal from the genuine signal.  We also use cosmic-rays on a  LaB$_6$ photocathode, which has higher than CsI's resistance to air and to ions drifting backwards onto the photocathode, to explore its quantum efficiency.
\end{abstract}

\section{Introduction and motivation}
The Belle II experiment is an hermetic magnetic spectrometer surrounded by particle-identification detectors and a CsI crystal calorimeter~\cite{TDC}. It studies billions of $B$, $D$, and $\tau$ particles from $e^+e^-$ collisions at the $\Upsilon$(4S) energy, produced by the KEK's SuperKEKB collider~\cite{Akai_2018}.
The calorimeter has a central in the Belle II physics program because its hermeticity, efficiency, and accurate detection performances enable unique important measurements. Beam-background photons degrade the performance of the calorimeter due to its proximity to the beams, technology, and large acceptance. These background photons are energy deposits unrelated with the $e^+e^-$ collision, but originating from single-beam interactions with residual gas in the vacuum~\cite{background}, induced by the tightly constrained collision environment demanded by the SuperKEKB nano-beam scheme~\cite{nanobeam}. They have 1--2 MeV typical energies, but can reach sufficiently high values to be reconstructed as calorimeter deposits. Since the collision time is known accurately from the accelerator radio-frequency, an efficient countermeasure against beam-backgrounds is to leverage signal timing to discern non-collision photons, which are preferentially out-of-time.
To explore this strategy, we are developing the gaseous photomultiplier (GasPM)~\cite{gaspm}, a resistive-plate chamber (RPC) paired with a photocathode, that is expected to detect photons with $\mathcal{O}(10)$~ps time resolution over large instrumented surfaces at affordable cost.
Combined with a radiator, the GasPM also operates as a Cherenkov detector offering competitive alternatives micro-channel-plate (MCP)-PMTs~\cite{top_mcp_pmt} for large area applications.

\begin{figure}[htbp]
    \centering
    \hspace*{-0.3cm} 
    \includegraphics[width=1.05\linewidth]{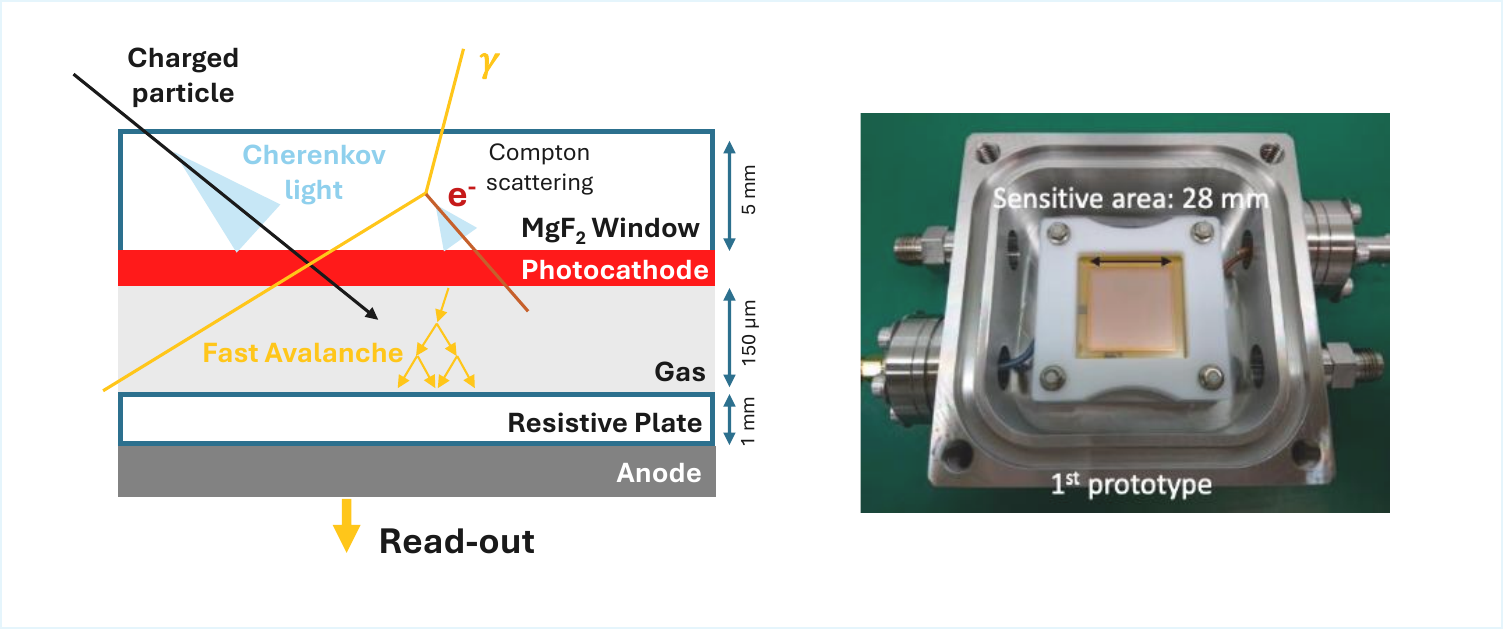}
    \caption{(Left) cross-sectional sketch of the GasPM prototype; (right) picture of a prototype (right).}
    \label{fig:gaspm_1st}
\end{figure}

\section{Design and operation principles}
Figure~\ref{fig:gaspm_1st} shows the GasPM architecture.  Photons traverse a transparent entry-window and impinge on the photocathode. In our application, $\mathcal{O}(1)$ MeV photons experience Compton scattering in the window. The outgoing electron emits Cherenkov photons that reach the photocathode. On the other hand, external incoming charged particles just emit Cherenkov photons in the window.
The constant electric field in the narrow gas-gap, filled with $90\%$ $\rm{C_2 H_2F_4}$ and $10\%$ $\rm{SF}_6$, accelerates the photoelectrons released from the photocathode generating an avalanche. The resulting charge drifts, inducing a signal at the copper electrode. The photocathode is installed on a transparent window with a soda-glass resistive plate inhibiting discharges. The window, photocathode, and resistive plate materials, bias voltage, gap thickness, gas admixture, can be adapted to one's specific application.

\begin{figure}[htbp]
    \centering
    \begin{minipage}{0.48\textwidth}
        \centering
        \includegraphics[width=\textwidth]{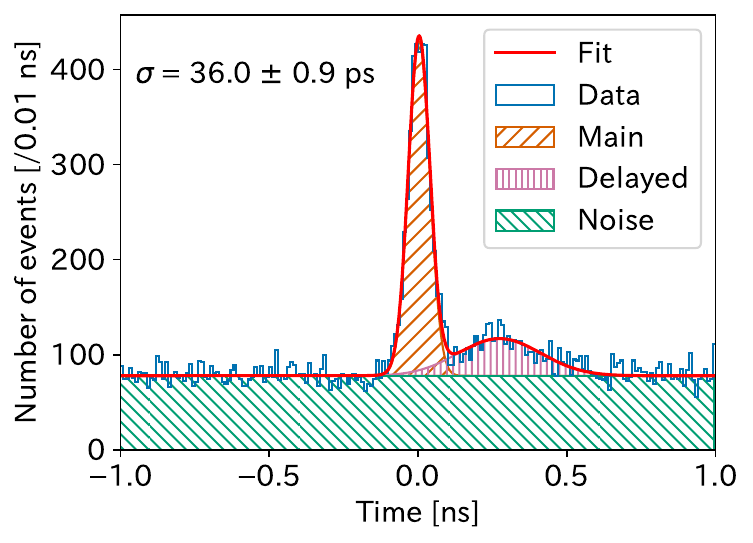}
    \end{minipage}
    \hfill
    \begin{minipage}{0.48\textwidth}
        \centering
        \includegraphics[width=0.9\textwidth]{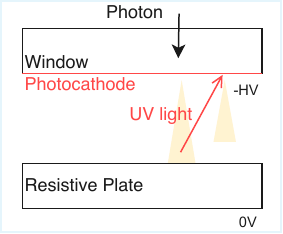}
    \end{minipage}
    \caption{(left) Distribution of single-photon time resolution obtained in a laser test~\cite{gaspm}; (right) simplified scheme of photon-feedback generation.}
    \label{fig:resol}
\end{figure}

\section{Previous work}
A laser test achieved good time resolution in 2022. Using a 375~nm picosecond-pulse laser; a quartz window; an LaB$_6$ photocathode, which has low quantum efficiency but stable performance when exposed to gases; a 170~$\mu$m gas gap for 176 kV/cm electric field; and a 1.1~mm thick TEMPAX float glass, we achieved $\sigma = \rm{25.0\pm1.1~ps}$ single-photon time resolution (Fig.~\ref{fig:resol}~(left))~\cite{gaspm}. We also observed a broad "shoulder", delayed by 0.1--0.5 ns.
In 2023 we demonstrated the Cherenkov application using a 3~GeV electron beam. We used a CsI photocathode, known to have the necessary UV quantum efficiency and sufficient tolerance to the relevant gas admixture; a 2.4~mm $\rm{MgF_2}$ window for UV transmission; and a  $200~\rm{\mu m}$ gas-gap with an electric field of just 140~kV/cm. We expected a worsening in time resolution due to reduced electrons-drift velocity.
Most signals exhibited overlapping pulses within, roughly, a  $600$~ns window. This made the timing determination difficult. The observed  resolution was $\sigma = 73.0\pm2.4~\rm{ps}$, exposing two main GasPM limitations. Photons emitted during the de-excitation of excited gas molecules hit the photocathode, generating secondary avalanches, so-called photon-feedback -- see Fig.~\ref{fig:resol}~(right). The resulting delayed signals overlap with the genuine primary signals, spoiling  time resolution. In addition, avalanche ions in the gas gap drift backwards and hit the photocathode, damaging it and thus degrading the efficiency over time (ion feedback)

\begin{figure}[htbp]
    \centering
    \includegraphics[width=\linewidth]{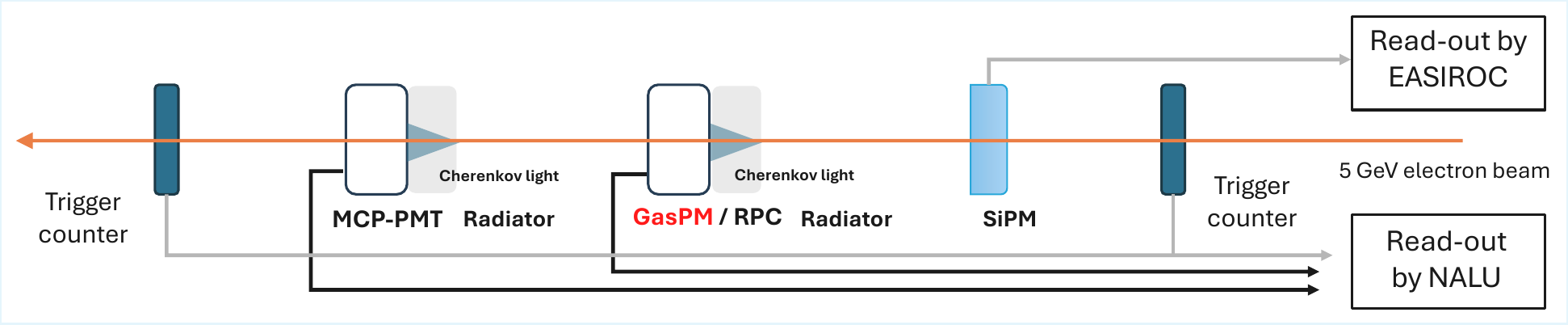}
    \caption{Beam-test setup scheme}
    \label{fig:beam_test}
\end{figure}

\section{Beam-test study of photon feedback}
We report on a test with  5~GeV electrons from the KEK PF-AR test beam-line on a GasPM prototype modified to improve the time resolution by understanding photon feedback. The electric field is  increased to 187~kV/cm by shrinking the gap to 150~$\mu$m; the MgF$_2$ window is thickened to 5~mm to increase the photon yield.
In addition, we double the read-out rate to 10~GSPS using the Nalu DSA-C10-8+ digitizer~\cite{NALU}. The objective is to discern primary from photon-feedback pulses. We also use  soda-glass resistive plate.
The experimental configuration, sketched in Figure~\ref{fig:beam_test}, includes an MCP-PMT with a quartz radiator as a time reference, a SiPM array with an acrylic radiator to veto $\delta$-ray–induced multi-electron events, and two plastic scintillators read out by PMTs as trigger counters  all along the beam line. See ref.~\cite{Garnero2026} for more details.

\begin{figure}[htbp]
    \centering
    \begin{minipage}{0.48\textwidth}
        \centering
        \includegraphics[width=\textwidth]{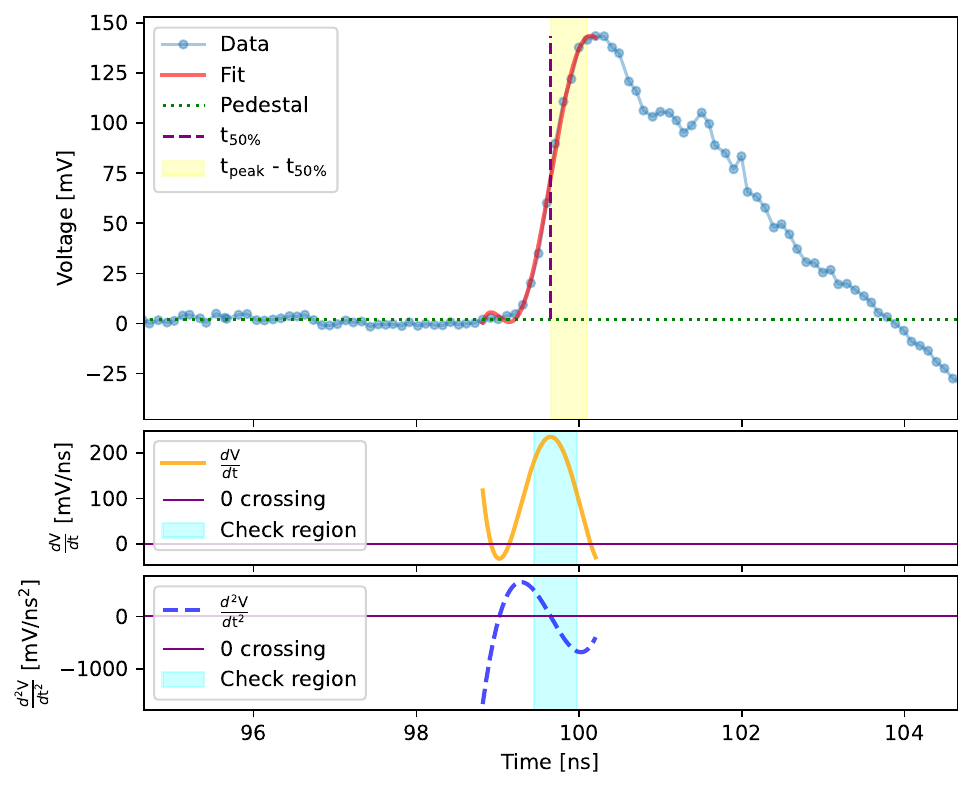}
    \end{minipage}
    \hfill
    \begin{minipage}{0.48\textwidth}
        \centering
        \includegraphics[width=\textwidth]{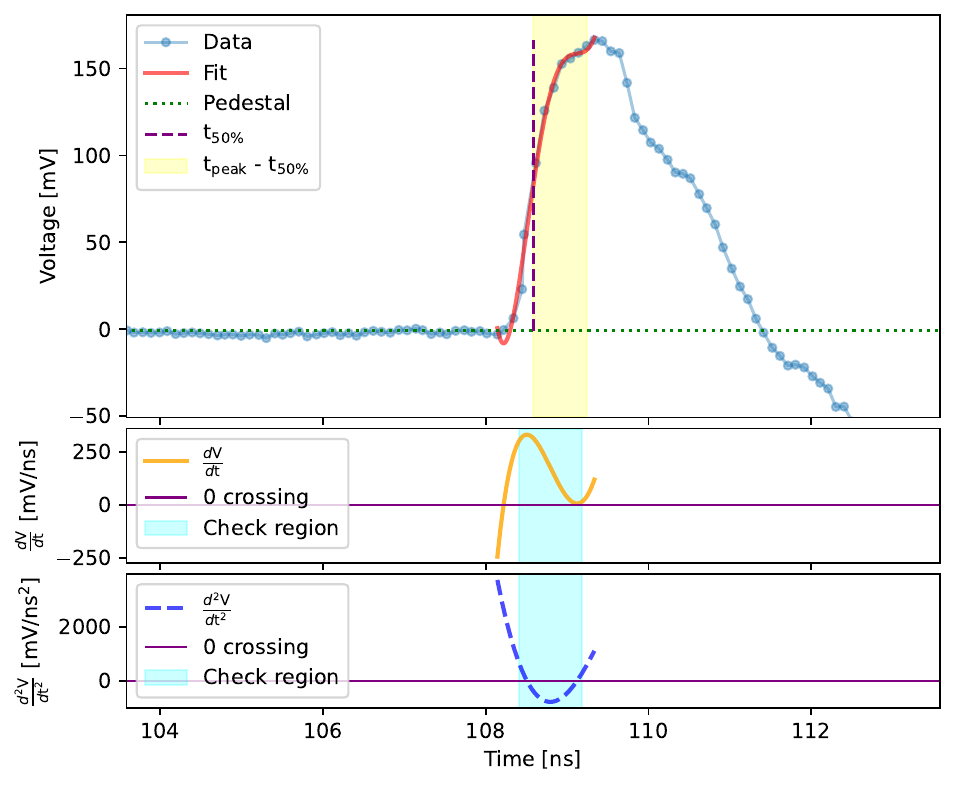}
    \end{minipage}
    \caption{Digitized waveforms from (left) beam-test event tagged as single avalanche and (right) event tagged as photon feedback beam-test with rising-edge fit projections overlaid. The relevant derivatives are shown in the bottom panels.}
    \label{fig:PF}
\end{figure}

\begin{figure}[htbp]
    \centering
    \includegraphics[width=0.6\linewidth]{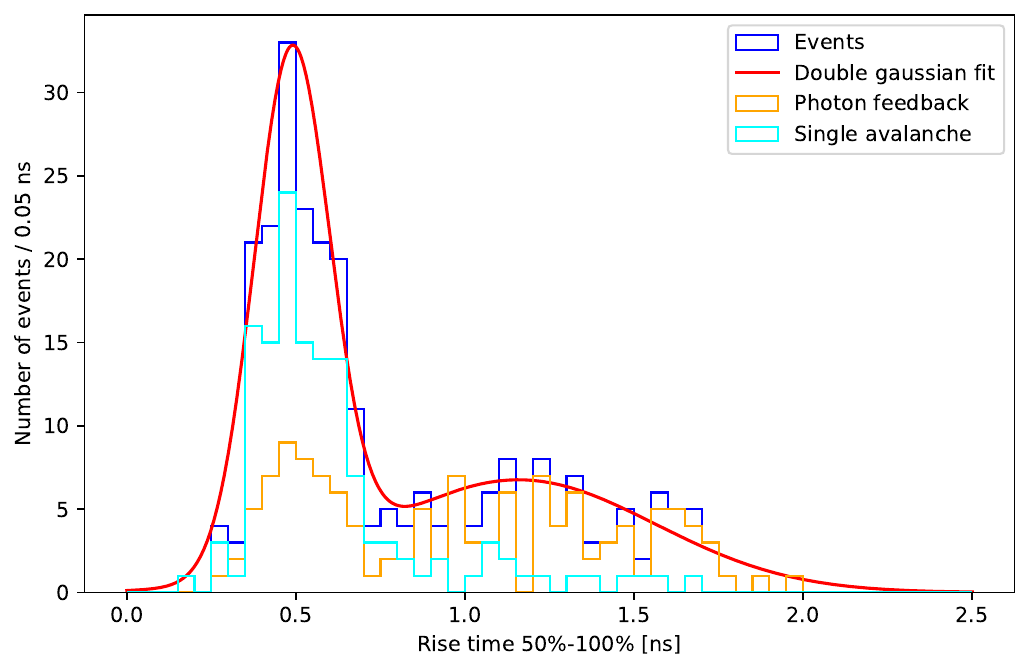}
    \caption{Signal rise time from 50\% to the peak after applying our photon-feedback identification approach. Cyan and orange curves indicate events tagged as single avalanche and photon-feedback, respectively.}
    \label{fig:PF_selection}
\end{figure}

Event classification is challenging as several processes occur simultaneously spoiling the correlation between pulse height, charge, and photon feedback, observed in the laser test. In addition, the thinner gas gap increases overlap between secondary and primary signals. Hence, 
we focus on the signal rising edge exploring an approach to discern statistically photon-feedback events.
The rising edge of the waveform are data sampled from the pedestal to peak, and fit it with a 8th-degree polynomial. We then restrict the analysis to events showing, in the second derivative of fitting function at 0.2 ns from the boundaries,two or more, zero-crossings. Multiple zero-crossings signal different curvatures, which are typically inconsistent with a single-component rise. The waveforms of two events are compared in Fig.~\ref{fig:PF}. Our approach identifies (53.2~$\pm$~2.3)$\%$ of events as those undergoing by photon feedback. The criterion is supported by a  50\%-to-100\% rise-time study (Fig.~\ref{fig:PF_selection}). 
Photon-feedback-identification is improved by thickening the gas gap, which extends the time-distance between the primary and the photon-feedback signal, facilitating  separation. Larger thickness, however, impacts gain. 

\section{Study of the LaB$_6$ photocathode}
We explore the LaB$_6$ photocathode under cosmic rays as a possible countermeasure against ion feedback. LaB6 is known to withstand better than CsI air exposure and ion feedback. However, in the 2022 laser test it showed poor quantum efficiency (QE) at 375~nm, although we expect better efficiency at Cherenkov wavelengths. Changes in the photocathode configuration are also introduced to improve UV performance.
We collect cosmic-ray data with both the GasPM and the RPC to subtract the ionisation contribution and study the Cherenkov signal.
The hit-rate is observed at (7.19~$\pm$~0.49)\% for the GasPM, industinguishible from the (7.66~$\pm$~0.18)\% RPC rate. This indicates that ionization-only signals largely dominate and therefore that the QE is too poor. An LED test attempts to assess the pure photon-detection rate. A clear signal-rate increase is observed between dark and illuminated conditions, confirming that the photocathode is responsive to light. However, the response remains low, confirming poor QE. Our conclusion is that  the LaB$_6$ photocathode fails to offer a viable option for the next beam test, at least in its current configuration.

\section{Summary}

We are developing the GasPM, as a  low-cost, large-coverage, $\mathcal{O}(10)$~ps-resolution RPC-based gaseous photodetector for mitigating beam background in a Belle II upgrade. Photon and ion feedback are significant  limitations. A beam test with a high-sampling-rate digitiser allows us to design an algorithm to suppress photon-feedback. Validation is ongoing along with an updated determination of time resolution. We also tested a LaB$_6$ photocathode, which shows poor QE at the target wavelength.


\section*{Acknowledgements}
This work was supported by DAIKO FOUNDATION and MEXT/JSPS KAKENHI Grant Numbers
JP26610068, JP16H00865, JP19H05099, JP21H01091, and JP23H05433. We acknowledge the support of KEK in our test at the PF-AR test beamline. S.G. acknowledges the IMAPP program and the SOKENDAI KEK Tsukuba/J-PARC Summer Student Program for the opportunity and funding.

\end{document}